\documentclass[prb,aps,twocolumn,floatfix,longbibliography,showpacs]{revtex4-1}
\usepackage{amssymb}
\usepackage{amsmath}
\usepackage{graphicx}
\usepackage{color}

\begin{document}

\title{Competition between the inter-valley scattering and the intra-valley scattering on magnetoconductivity induced by screened Coulomb disorder in Weyl semimetals}

\author{Xuan-Ting Ji$^{1}$}

\author{Hai-Zhou Lu$^{2}$}
\email{luhz@sustc.edu.cn}

\author{Zhen-Gang Zhu$^{1,3}$}
\email{zgzhu@ucas.ac.cn}

\author{Gang Su $^{3,4}$}
\email{gsu@ucas.ac.cn}

\affiliation{$^{1}$School of Electronic, Electrical and Communication Engineering, University of Chinese Academy of
Sciences, Beijing 100049, China. \\
$^{2}$Institute for Quantum Science and Engineering and Department of Physics, South University of Science and Technology of China, Shenzhen, 518055 China.\\
$^{3}$Theoretical Condensed Matter Physics and Computational Materials Physics Laboratory, School of Physical Sciences, University of Chinese Academy of Sciences, Beijing 100049, China.\\
$^{4}$Kavli Institute of Theoretical Sciences, University of Chinese Academy of Sciences, Beijing 100190, China.
}

%\date{\today }
\begin{abstract}
Recent experiments on Weyl semimetals reveal that charged impurities may play an important role. We use a screened Coulomb disorder to model the charged impurities, and study the magneto-transport in a two-node Weyl semimetal. It is found that when the external magnetic field is applied parallel to the electric field, the calculated longitudinal magnetoconductivity shows positive in the magnetic field, which is just the negative longitudinal magnetoresistivity (LMR) observed in experiments.  When the two fields are perpendicular to each other, the transverse magnetoconductivities are measured. It is found that the longitudinal (transverse) magnetoconductivity is suppressed (enhanced) sensitively with increasing the screening length. This feature makes it hardly to observe the negative LMR in Weyl semimetals experimentally owing to a small screening length. Our findings gain insight into further understanding on recently actively debated magneto-transport behaviors in Weyl semimetals. Furthermore we studied the relative weight of the inter-valley scattering and the intra-valley scattering. It shows that the former is as important as the latter and even dominates in the case of strong magnetic fields and small screening length. We emphasize that the discussions on inter-valley scattering is out of the realm of one-node model which has been studied.
\end{abstract}


\maketitle
% \tableofcontents

\section{Introduction}
% ---------------------------------------------------------------------------
Weyl fermion proposed by Hermann Weyl in 1929 as the massless solution of the Dirac equation is of great significance, and has been searched for several decades from the aspect of  particle physics. Neutrino has deemed to be Weyl fermion for a long time before the discovery of the neutrino oscillation\cite{PONTECORVO}, which suggests neutrino is massive. It still has null result in the standard model of the particle physics. Nevertheless, it has been suggested that low energy excitations in some particular condensed matters can be the solutions to the Weyl Hamiltonian, which was understood as Weyl semimetal \cite{Wan}.
Weyl semimetal soon becomes one of central issues of the topological materials \cite{Autes,Bulmash,Xu,Pavan Hosur1,Pavan Hosur2,Pavan Hosur3,Goswami,Ojanen}. Very recently, with the help of angle-resolved photoemission spectroscopy (ARPES), Weyl nodes have been identified in the TaAs family \cite{weng,B.Q. Lv,s huang,SXu}. Another candidate material with Weyl nodes is proposed to be YbMnBi$_{2}$\cite{Borisenko}. Moreover the second class of Weyl semimetal has been discorvered in MoTe$_{2}$ \cite{Deng}.

Weyl nodes are the band crossings and behave as monopoles due to the Berry curvature in the three-dimensional (3D) momentum space. The Weyl nodes appear in pairs with opposite chirality. They can only eliminate each other when meet together. Although the Weyl nodes are topologically robust, a chiral charge from one Weyl node can be pumped to the other in the presence of a parallel magnetic and electric field ($\mathbf{E}\cdot\mathbf{B}$ term), thus violating the total chirality. This was named as chiral anomaly \cite{Ninomiya}. It has been shown that chiral anomaly may induce diverse effects, such as frequency shift on the plasmon mode \cite{Zhou}, anomalous Hall effect \cite{Wan,Xu,K.Y.Yang,Burkov2,Yang} and chiral magnetic effect \cite{Kharzeev,Yamamoto,Stephanov,Landsteiner,Min-Fong Yan,G.Basar}, etc. Particularly it may manifests itself by a negative longitudinal magnetoresistance (LMR) which has been theoretically studied \cite{Son,Burkov,Shovkovy} and has been observed experimentally in the TaAs family \cite{Xiaochun Huang,Zhang:2015gwa,Zhangc,arnold,Yangx,Wangz}.

Theoretically, a semiclassical mechanism of the LMR was introduced, wherein the charge transport is dominated by guiding center motion \cite{Song}. Pesin \textit{et. al.} studied the long-range-disorder-affected density of states in absence of or in presence of weak magnetic field and found that a finite density of states can be induced by the long-range potential fluctuations as the energy approaching the degeneracy point \cite{Pesin}. Klier \textit{et. al.} \cite{Klier} explored the density of states, Landau levels broadening, and the magnetoresistivity in transversal magnetic fields within short-range impurities or charged (Coulomb) impurities. Moreover, $\delta$-potential disorder and random-Gaussian-potential disorder in Weyl semimetal near the Weyl nodes have been studied \cite{Zhang,Lu1}.

Although there are intensive investigations on the theory aspect, it is still insufficient to understand the realistic features observed in experiments because most of previous studies treated one-node model. As known the nodes exist in pairs and the chiral anomaly involves at least one pair. Therefore it is not possible to explore the chiral anomaly in the one-node model. Furthermore, it is noted that the observed LMR in some experiments is not negative but positive; while it is negative in some others.
And the Fermi level varies dramatically in different samples. These sensitively sample-dependent features indicate that the charged impurities may be the main type of disorder in the transport measurements. %Although $\delta$-potential disorder and random-Gaussian-potential disorder in Weyl semimetal near the Weyl nodes have been studied \cite{Zhang,Lu1},  it is not sufficient for the understanding of the observed features due to the ignorance of the inter-valley scattering.
For charged impurities the Coulomb interaction between them and their surrounding electrons is dressed, leading to a screened Coulomb interaction, which induces both intra-valley scattering (when we describe the scattering, we use "valley" rather "node" for convenience) and inter-valley scattering. Therefore, to study the effect of the inter-valley scattering we need start from at least a two-node model.

In this paper, we studied such a two-node model. The inter-valley scattering and the intra-valley scattering are considered simultaneously. We show a negative LMR in magnetic field, consistent with the prediction of the chiral anomaly \cite{Zhang}. It is seen that the transverse magnetoconductivity is enhanced sensitively by increasing the screening length which is just opposite to that of longitudinal magnetoconductivity. Based on this result, we can understand recent observations in experiments and find out a way to facilitate the observation of the negative LMR in the future experiments. We also show that the inter-valley scattering can be as important as the intra-valley scattering, or even more important than the latter. The results on the inter-valley scattering are novel since previously one-node model was mainly explored in which the inter-valley scattering can not be cast.

The paper is organized as follows. We first introduce the two-node model in Sec. II. We then discuss the
formulas and analysis of the longitudinal and transverse magnetoconductivities in the presence of the Coulomb screening
disorder in Sec. III. The conclusions are given in Sec. IV.

\section{Model}
\subsection{$\mathbf{B}=0$ case}
In a two-node model, the energy dispersion is linear in momentum around the nodes forming cones (or valley), which are labeled by $K$ cone and $K^{\prime}$ cone. The two-node Hamiltonian \cite{Lu2} in absence of an external magnetic field $\mathbf{B}$ is then given
\begin{equation}
\mathcal{H}=\nu\hbar v_{F}\left[\left(\mathbf{k}+\nu\mathbf{k}_{c}\right)\cdot\mathbf{\boldsymbol{\sigma}}\right],
\end{equation}
where $\nu=1$ for $K$ cone and $\nu=-1$ for $K^{\prime}$ cone, $v_{F}$ is the Fermi velocity and $\mathbf{k}_{c}=\left(0,0,\pm k_{c}\right)$ represent the positions of the two Weyl nodes in the momentum space.
For convenience, we explicitly re-express this Hamiltonian separately,
\begin{eqnarray}
\mathcal{H}_{K}&=&\hbar v_{F}\left(\mathbf{k}+\mathbf{k}_{c}\right)\cdot\boldsymbol{\sigma} ,\nonumber\\
\mathcal{H}_{K^{\prime}}&=&-\hbar v_{F}\left(\mathbf{k}-\mathbf{k}_{c}\right)\cdot\boldsymbol{\sigma}.
\label{original-Hamiltonian}
\end{eqnarray}
By a translation, we get two Hamiltonians which only depend on the relative momentum $\mathbf{k}$ with respect to the two nodes as
\begin{eqnarray}
\tilde{H}_{K}&=&\mathcal{H}_{K}-\hbar v_{F}\mathbf{k}_{c}\cdot\boldsymbol\sigma,\nonumber\\
\tilde{H}_{K^{\prime}}&=&\mathcal{H}_{K^{\prime}}-\hbar v_{F}\mathbf{k}_{c}\cdot\boldsymbol\sigma.  \label{Hamiltonian}
\end{eqnarray}
These two Hamiltonians satisfy the time reversal symmetry (TRS).

For $K$ cone, we have eigenstates of the Hamiltonian in Eq. (\ref{Hamiltonian})
\begin{eqnarray}
\Psi_{+K}=\left(\begin{array}{c}
\cos\frac{\theta}{2}\\
\sin\frac{\theta}{2}e^{i\varphi}
\end{array}\right),
\Psi_{-K}=\left(\begin{array}{c}
\sin\frac{\theta}{2}\\
-\cos\frac{\theta}{2}e^{i\varphi}
\end{array}\right),
\end{eqnarray}
where $\pm$ represent the conduction band and valence band, respectively.
The eigenenergies are thus
%\begin{eqnarray}
%\tilde{E}_{+K}&=&\hbar v_{F}\sqrt{k_{x}^{2}+k_{y}^{2}+k_{z}^{2}} ,\nonumber\\
%\tilde{E}_{-K}&=&-\hbar v_{F}\sqrt{k_{x}^{2}+k_{y}^{2}+k_{z}^{2}} .
%\end{eqnarray}
\begin{equation}
\tilde{E}_{\pm K}=\pm\hbar v_{F}\sqrt{k_{x}^{2}+k_{y}^{2}+k_{z}^{2}} ,
%\tilde{E}_{-K}&=&-\hbar v_{F}\sqrt{k_{x}^{2}+k_{y}^{2}+k_{z}^{2}} .
\end{equation}
The angle $\theta$ and $\varphi$ are decided by the following relations
\begin{eqnarray}
\cos\theta=\frac{k_{z}}{|\tilde{E}|},
\tan\varphi=\frac{k_{x}}{k_{y}}.
\end{eqnarray}
The eigenenergies of the original Hamiltonian in Eq. (\ref{original-Hamiltonian}) are given by
\begin{eqnarray}
E_{K}=\tilde{E}_{\pm K}+\hbar v_{F}k_{c}.
\end{eqnarray}
In $K^{\prime}$ cone, the relative momentum is expressed by $\mathbf{k}^{\prime}$ and satisfies the relations, $\mathbf{k^{\prime}}$=-$\mathbf{k}$, $\theta^{\prime}$=$\pi-\theta$, $\varphi^{\prime}$=$\pi+\varphi$ due to the TRS.
We therefore get the eigenstates for the $K^{\prime}$ cone
\begin{eqnarray}
\Psi_{+K^{\prime}}=\left(\begin{array}{c}
\sin\frac{\theta}{2}\\
-\cos\frac{\theta}{2}e^{i\varphi}
\end{array}\right),
\Psi_{-K^{\prime}}=\left(\begin{array}{c}
\cos\frac{\theta}{2}\\
\sin\frac{\theta}{2}e^{i\varphi}
\end{array}\right).
\end{eqnarray}
The corresponding eigenenergies for $K^{\prime}$ cone are
\begin{eqnarray}
E_{K^{\prime}}=\tilde{E}_{\pm K^{\prime}}+\hbar v_{F}k_{c}.
\end{eqnarray}
where
\begin{eqnarray}
\tilde{E}_{\pm K^{\prime}}=\mp\hbar v_{F}\sqrt{k_{x}^{2}+k_{y}^{2}+k_{z}^{2}}.
\end{eqnarray}

\subsection{$\mathbf{B}\neq0$ case}
When a magnetic field $\mathbf{B}=(0,0,B)$ is applied along the z direction, Landau bands form and disperse with $k_{z}$. Here we choose the Landau gauge $\mathbf{A}=(-yB,0,0)$. Then we have
\begin{eqnarray}
\mathbf{k}=\left(k_{x}-\frac{eB}{\hbar}y,-i\partial_{y},k_{z}\right).
\end{eqnarray}
Defining the ladder operators,
\begin{eqnarray}
a&=&-[(y-l_{B}^{2}k_{x})/l_{B}+l_{B}\partial_{y}]/\sqrt{2},\nonumber\\
a^{\dagger}&=&-[(y-l_{B}^{2}k_{x})/l_{B}-l_{B}\partial_{y}]/\sqrt{2},
\end{eqnarray}
where $l_{B}=\sqrt{\hbar/eB}$ is the magnetic length. For $K$ cone, we have
\begin{eqnarray}
\tilde{H}_{K}=\hbar v_{F}\left(\begin{array}{cc}
k_{z} & -\frac{\sqrt{2}}{l_{B}}a\\
-\frac{\sqrt{2}}{l_{B}}a^{\dagger} & -k_{z}
\end{array}\right). \label{Magneto Hamiltonian}
\end{eqnarray}
The corresponding eigenstates of the Landau bands are given by
\begin{eqnarray}
|n\geqslant1,k_{x},k_{z},+K\rangle&=&\left[\begin{array}{c}
\cos\frac{\theta_{n}^{k_{z}}}{2}|n-1\rangle\\
\sin\frac{\theta_{n}^{k_{z}}}{2}|n\rangle
\end{array}\right]|k_{x},k_{z}\rangle , \nonumber\\
|n\geqslant1,k_{x},k_{z},-K\rangle&=&\left[\begin{array}{c}
\sin\frac{\theta_{n}^{k_{z}}}{2}|n-1\rangle\\
-\cos\frac{\theta_{n}^{k_{z}}}{2}|n\rangle
\end{array}\right]|k_{x},k_{z}\rangle,
\label{EigenstatesK}
\end{eqnarray}
where $\cos\theta_{n}^{k_{z}}=\frac{k_{z}}{\sqrt{k_{z}^{2}+2n/l_{B}^{2}}}$, $n$ is the index of the Landau bands. Please note that $\theta_{n}^{k_{z}}$ is the same in $K$ and $K'$ cone, and the TRS has been taken into account already.
The eigenenergies for the Hamiltonian (\ref{Magneto Hamiltonian}) are
\begin{eqnarray}
\tilde{E}_{\pm K}^{n} =\pm\hbar v_{F}\sqrt{k_{z}^{2}+2n/l_{B}^{2}}.
\end{eqnarray}
The eigenenergies for the original Hamiltonian are
\begin{eqnarray}
{E}_{\pm K}^{n}=\tilde{E}_{\pm K}^{n}+\hbar v_{F}k_{c}.
\end{eqnarray}
The eigenstate of the $n=0$ Landau band is obtained
\begin{eqnarray}
|n=0,k_{x},k_{z},K\rangle=\left[\begin{array}{c}
0\\
|0\rangle
\end{array}\right]|k_{x},k_{z}\rangle,
\label{Eigen0K}
\end{eqnarray}
with the eigenenergy
\begin{eqnarray}
E_{k_{z}}^{0K}=-\hbar v_{F}k_{z}+\hbar v_{F}k_{c}.
\end{eqnarray}
The definition of $\theta^{k_{z}}_{0K}$ is given
\begin{eqnarray}
\cos\left(\theta^{k_{z}}_{0K}/2\right) &=& \sqrt{\left(1+k_{z}/\sqrt{k_{z}^{2}}\right)/2}=\Theta(k_{z}), \notag \\
\sin\left(\theta^{k_{z}}_{0K}/2\right) &=& \sqrt{\left(1-k_{z}/\sqrt{k_{z}^{2}}\right)/2}=\Theta(-k_{z}),
\label{theta0K}
\end{eqnarray}
where $\Theta(x)$ is the step-function.

Similarly, the eigenstates for the $K^{\prime}$ cone are given by
\begin{eqnarray}
|n\geqslant1,k_{x},k_{z},+K^{\prime}\rangle&=&\left[\begin{array}{c}
\sin\frac{\theta_{n}^{k_{z}}}{2}|n-1\rangle\\
-\cos\frac{\theta_{n}^{k_{z}}}{2}|n\rangle
\end{array}\right]|k_{x},k_{z}\rangle  , \nonumber\\
|n\geqslant1,k_{x},k_{z},-K^{\prime}\rangle&=&\left[\begin{array}{c}
\cos\frac{\theta_{n}^{k_{z}}}{2}|n-1\rangle\\
\sin\frac{\theta_{n}^{k_{z}}}{2}|n\rangle
\end{array}\right]|k_{x},k_{z}\rangle.
\label{EigenstatesKprime}
\end{eqnarray}
The corresponding eigenenergies are obtained
\begin{eqnarray}
\tilde{E}_{\pm K^{\prime}}^{n}&=&\mp\hbar v_{F}\sqrt{k_{z}^{2}+2n/l_{B}^{2}} , \nonumber\\
E_{\pm K^{\prime}}^{n}&=&\tilde{E}_{\pm K^{\prime}}^{n}+\hbar v_{F}k_{c} .
\end{eqnarray}
The eigenstate of the $n=0$ Landau band is obtained
\begin{eqnarray}
|n=0,k_{x},k_{z},K^{\prime}\rangle=\left[\begin{array}{c}
|0\rangle\\
0
\end{array}\right]|k_{x},k_{z}\rangle,
\label{Eigen0Kprime}
\end{eqnarray}
with the eigenenergy
\begin{eqnarray}
E_{k_{z}}^{0K^{\prime}}=\hbar v_{F}k_{z}+\hbar v_{F}k_{c}.
\end{eqnarray}
According to the TRS, $\theta^{k_{z}}_{0K'}=\pi-\theta^{k_{z}}_{0K}$, hence we have $\cos(\theta^{k_{z}}_{0K'}/2)=\sin(\theta^{k_{z}}_{0K}/2)$ and $\sin(\theta^{k_{z}}_{0K'}/2)=\cos(\theta^{k_{z}}_{0K}/2)$.

\section{Formalism and Results}
\subsection{Longitudinal Magnetoconductivity}
In the case of low temperature and a Fermi level close to the Weyl nodes,
the transport is essentially involved with the $0$-th Landau band.
When the applied electric and magnetic fields are parallel, the changing rate of density of charge carriers near one
node is maximal due to the chiral anomaly. In this
case, the semiclassical conductivity of the $0$-th Landau band can be derived in terms of
the standard Green's function method. Alternatively,
it can be simply figured out by using the Einstein relation $\sigma_{zz}=e^{2}N_{F}D,$
where the density of state (DOS) can be found as the Landau degeneracy
times the DOS of one-dimensional systems, i.e., $N_{F}=(1/2\pi l_{B}^{2})\times(2/\pi\hbar v_{F})$.
$D=v_{F}^{2}\tau^{0,\text{tr}}_{k_F}$ is the diffusion coefficient.  $\tau^{0,\text{tr}}_{k_F}$ is the transport time, and $k_F$ is the Fermi wave number. Under the first-order Born approximation, for the scattering among the states on the Fermi surface of the $0$-th band, the transport time can be obtained
\begin{eqnarray}\label{transport time1}
\frac{\hbar}{\tau^{0,\text{tr}}_{k_z}}
&=& 2\pi \sum_{k_x',k_z'}
\langle |U^{0,0}_{k_x,k_z;k_x',k_z'}|^2 \rangle \Lambda\delta(E_F-E^0_{k_z'})\nonumber\\
&&\times \left(1-\frac{v^z_{0,k_z'}}{v_F}\right),
\end{eqnarray}
where $U^{0,0}_{k_x,k_z;k_x',k_z'}$ represents the scattering matrix elements and $\langle ... \rangle$ means the impurity average. An extra factor $\Lambda$ is introduced to correct the unphysical van Hove singularity near the band edge \cite{Songbo}, and the details of $\Lambda$ can be found in the Appendix \ref{lambda}. We note that, for z-axis magnetic field, only the diagonal elements of the velocity matrix with respect to the z-direction contribute to the longitudinal magnetoconductivity. On the contrary, only the off-diagonal elements of the velocity matrix with respect to the x-direction contribute to the transverse magnetoconductivity. %When the Fermi energy lies at the vicinity of the Weyl nodes, transport time is the unique component to be considered.

The transport time $\tau^{0,\text{tr}}_{k_z}$ is sensitive to the scattering potential. To reflect the realistic situation, we use the screened Coulomb potential to model the charged impurities,
\begin{eqnarray}
\phi(r)=\frac{e^{2}}{\alpha\kappa_{0}r}e^{-\frac{r}{\lambda}},
\end{eqnarray}
where $\lambda$ measures the screening radius, which can be decided by a finite chemical potential $\mu$, temperature $T$ and the excess electron density $n(\mu,T)$ according to the Eq. (11) in Ref. \onlinecite{Rodionov}. The $n(\mu,T)$ is defined by
\begin{eqnarray}
n_{D}-n_{A}=n(\mu,T),
\end{eqnarray}
where $n_{D}$, and $n_{A}$ denote the density of the donors and acceptors, respectively. $\alpha\kappa_{0}$ represents for the dielectric constant, $\alpha$ is a dimensionless constant and $e$ is the charge of the electron.

The longitudinal conductivity $\sigma_{zz}$ can be found from the formula
\begin{eqnarray}\sigma_{zz}&=&\frac{e^{2}\hbar}{2\pi V}\sum_{k_{z},k_{x}}\text{Tr}\left(\mathbf{v}_{0,k_{z}}^{z}\mathbf{G}_{0,k_{z}}^{R}\mathbf{\widetilde{v}}_{0,k_{z}}^{z}\mathbf{G}_{0,k_{z}}^{A}\right),
\end{eqnarray}
where bold symbols, e.g. $\mathbf{v}$ and $\mathbf{G}$, are all diagonal matrixes in the valley space (i.e. cone space), and the subscript $0$ represents the $n=0$ Landau subspace. We thus have
\begin{equation}
\sigma_{zz}=\frac{e^{2}\hbar}{2\pi V}\underset{k_{x},k_{z},i}{\sum}\left( v_{0,k_{z}}^{z,i}\tilde{v}_{0,k_{z}}^{z,i}G_{0i}^{R}G_{0i}^{A}\right),
\end{equation}
where $i$ runs over $K$ and $K^{\prime}$ cone.
Note that, the electrons in different cone move in opposite directions due to the chirality. In our case, we fixed the fermi energy at the zeroth Landau band, and all the Landau bands in the $K^{\prime}$ cone
are fully occupied. Hence, we have
\begin{eqnarray}
\sigma_{zz}=\frac{e^{2}\hbar}{2\pi V}\underset{k_{x},k_{z}}{\sum}\left(v_{0,k_{z}}^{zK}\tilde{v}_{0,k_{z}}^{zK}G_{0K}^{R}G_{0K}^{A}\right),
\end{eqnarray}
where
\begin{eqnarray}
v_{0,k_z}^{zK}&=&\partial E_{k_{z}}^{0K}/\hbar\partial k_{z}=-v_{F},
\end{eqnarray}
is the $\hat{\mathbf{z}}$-direction velocity of an electron in the $0$-th band. $\widetilde{v}_{0,k_z}^{zK}$ is the dressed velocity after taking into account the vertex correction. The retarded/advanced Green's function for the $0$-th band are $G_{0K}^{R/A}=1/(E_{F}-E_{k_{z}}^{0K}\pm i\hbar/2\tau)$ where $\tau$ is the corresponding momentum relaxation time which is related to the transition probability (induced by the impurities) in $K$ cone at $n=0$ band. $V=L_{x}L_{y}L_{z}$ is the volume of a box.

%%%%%%%%%%%%%%%%%%%%%%%%%%%%%%%%%%%%%%%%%%%%%%%%%%%%%%%%%%%%%%%%%%%%%%%%%%%%%%%%%%%%%%%%%%%%%%%%%%%%%%%%%%%%%%%%%%%%%%%%%%%%%%%%%%%%%%%%%
\begin{figure}[tb]
\includegraphics[width=0.85\linewidth]{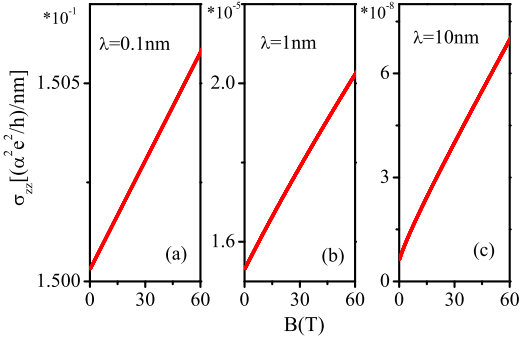}
%\vspace{-0.5cm}
\caption[]{The longitudinal magnetoconductivity versus the magnetic flux density for different screening lengths. Here $k_{c}=0.1$ nm$^{-1}$.}\label{sigzz}
\end{figure}
%%%%%%%%%%%%%%%%%%%%%%%%%%%%%%%%%%%%%%%%%%%%%%%%%%%%%%%%%%%%%%%%%%%%%%%%%%%%%%%%%%%%%%%%%%%%%%%%%%%%%%%%%%%%%%%%%%%%%%%%%%%%%%%%%%%%%%%%%

In the diffusive regime, one can replace $G^{R}_{0K}\widetilde{v}_{0,k_z}^{zK}G^{A}_{0K}$ by $(2\pi/\hbar)\tau^{0K,\text{tr}}_{k_z} v_{0,k_z}^{zK} \Lambda\delta(E_F-E^{0K}_{k_z})$. We get
\begin{equation}
\sigma_{zz}=\frac{e^{2}}{V}\sum_{k_{z},k_{x}}(v_{0,k_{z}}^{zK})^{2}\tau_{k_{F}}^{0K,\text{tr}}\Lambda\delta(E_{F}-E_{k_{z}}^{0K}).
\label{sigmazz-1}
\end{equation}
%
%{\color{blue}{The detailed derivation of Eq. (\ref{sigmazz-1}) can be found in Appendix \ref{zztransport}.}}
We can expand the Delta function as $\delta(E_F-E^{0K}_{k_z})=\delta(k_{F}+k_{z}-k_{c})/\hbar v_F$ since the equation, i.e. $E_{F}-E_{k_{z}}^{0K}=0$, leads us only one root at $k_{F}=k_{c}-k_{z}$.
Similarly, expanding the Delta function as $\delta(E_F-E^{0K^{\prime}}_{k_z})=\delta(k_{F}-k_{z}-k_{c})/\hbar v_F$ due to $E_{F}-E_{k_{z}}^{0K^{\prime}}=0$, leads us only one root at $k_{F}=k_{c}+k_{z}$.
 This allows us to perform the summation over $k_z$. And we have
\begin{eqnarray}
\sigma_{zz}&=&\frac{e^{2}}{V}\underset{k_{x,}k_{z}}{\sum}\left(v_{0,k_{z}}^{zK}\right)^{2}\Lambda\tau_{k_{F}}^{0K,\text{tr}}\delta(k_{F}+k_{z}-k_{c})/\hbar v_{F}\nonumber\\
&=&\frac{L_{z}e^{2}}{V\hbar v_{F}}\underset{k_{x}}{\sum}\int\frac{dk_{z}}{2\pi}\left(v_{0,k_{z}}^{zK}\right)^{2}\Lambda\tau_{k_{F}}^{0K,\text{tr}}\delta(k_{F}+k_{z}-k_{c})\nonumber\\
&=& \frac{e^2}{h} \frac{v_{F}}{L_xL_y} \sum_{k_x}  \tau^{0K,\text{tr}}_{k_F} \Lambda.
\end{eqnarray}
We need to calculate the transport time $\tau^{0K,\text{tr}}_{k_F}$ at the Fermi surface, which can be found in the Appendix \ref{zztransport}. After a straight forward calculation, we obtain
\begin{eqnarray}
\sigma_{zz}
& = & \frac{e^{2}}{h}
\frac{(\hbar v_F)^2 }{2\pi n_{\text{imp}}\left(\frac{4\pi e^{2}}{\alpha\kappa_{0}}\right)^{2}I_{1}l_B^4}.\label{sigmazzf}
\end{eqnarray}

In Fig.~\ref{sigzz}, we plot the longitudinal magnetoconductivity versus $B$ at different screening length $\lambda$. For clarity, we treat the $\lambda$ as a varying parameter and focus on investigation its effect on the relative weight of the inter-valley scattering to the intra-valley scattering. It is noted that $\sigma_{zz}$ is linear in $B$. This linear dependence on $B$ was also obtained in Refs. \onlinecite{Son,Zhang,Shovkovy}. However the effect of screened Coulomb disorder on $\sigma_{zz}$ is unknown so far.
It shows that the magnitude of $\sigma_{zz}$ is inversely proportional to $\lambda^{4}$ when $\lambda$ is small.
This $\lambda^4$-dependence results from the $\lambda$ dependence of $I_{1}$. In other words, a larger $\lambda$ induces a larger longitudinal magnetoresistance. This result seems in contradiction to our intuitive expectation that a small $\lambda$ may induce a stronger inter-valley scattering leading to a larger LMR due to Heisenberg uncertainty principle. However, this effect may be overwhelmed by another effect, i.e. a weak scattering potential for small $\lambda$. The overall behavior stems from that the screened Coulomb potential induces not only the intra-valley scattering but also the inter-valley scattering. We should emphasize that the screening length $\lambda$ in our calculation is taken from $0.1$ nm to $10$ nm. This interval is consistent with that in Ref. \onlinecite{Klier} and Ref. \onlinecite{Rodionov}.

\subsection{Transverse magnetoconductivity \label{Sec:Sxx}}
After displaying the results for longitudinal magneto-conductivity, we study in this subsection the transverse magnetoconductivity, i.e. $\sigma_{xx}$. For consistency, the Fermi energy is assumed to be near the Weyl nodes, and the conductivity along the $\hat{\bf{x}}$ direction can be written as
\begin{eqnarray}\label{sigma-xx}
\sigma_{xx}=\frac{e^{2}\hbar}{2\pi V}\underset{\mathbf{k}}{\sum}\text{Tr}\left(\mathbf{G}^{R}\mathbf{v}_{\mathbf{k}}^{x}\mathbf{G}^{A}\mathbf{v}_{\mathbf{k}}^{x}\right).
\end{eqnarray}
$\mathbf{G}$ and $\mathbf{v}$ are all matrixes and we have
%\begin{widetext}
\begin{equation}
\sigma_{xx}=\frac{e^{2}\hbar}{\pi V}\underset{k_{x,}k_{z},i,i'}{\sum}\Re\left(G_{0i}^{R}v_{0i,1+i'}^{x}G_{1+i'}^{A}v_{1+i',0i}^{x}\right),
\label{sigma-xx1}
\end{equation}
%\end{widetext}
where $i,i'$ run over $K$ and $K'$ indexes. A derivation of Eq. (\ref{sigma-xx1}) is given in Appendix \ref{matrixformsigmaxx}.  As an example, $v_{0K,1+K}^{x}$ links a scattering from zeroth band in $K$ cone to the first Landau band in $K$ cone.

With the definition of the velocity $v^{x}=\frac{1}{\hbar}\frac{\partial H}{\partial k_{x}}=\nu v_{F}\sigma_{x}$, the elements are derived
\begin{eqnarray}
v_{0K,1+K}^{x}&=&v_{F}\sin\frac{\theta_{0}^{k_{z}}}{2}\cos\frac{\theta_{1}^{k_{z}}}{2}=v_{F}\Theta(-k_{z})\cos\frac{\theta_{1}^{k_{z}}}{2} ,\nonumber\\
v_{0K,1+K^{\prime}}^{x}&=&v_{F}\sin\frac{\theta_{0}^{k_{z}}}{2}\sin\frac{\theta_{1}^{k_{z}}}{2}=v_{F}\Theta(-k_{z})\sin\frac{\theta_{1}^{k_{z}}}{2} ,\nonumber\\
v_{0K^{\prime},1+K}^{x}&=&v_{F}\cos\frac{\theta_{0}^{k_{z}}}{2}\cos\frac{\theta_{1}^{k_{z}}}{2}=v_{F}\Theta(-k_{z})\cos\frac{\theta_{1}^{k_{z}}}{2},\nonumber\\
v_{0K^{\prime},1+K^{\prime}}^{x}&=&v_{F}\cos\frac{\theta_{0}^{k_{z}}}{2}\sin\frac{\theta_{1}^{k_{z}}}{2}=v_{F}\Theta(-k_{z})\sin\frac{\theta_{1}^{k_{z}}}{2}.\nonumber\\
\end{eqnarray}
The momentum relaxation time for $i$ valley ($i=K, K'$) is defined as
\begin{equation}
\frac{1}{\tau_{i}}\equiv\frac{1}{\tau_{0i}}+\frac{1}{\tau_{Ii}},
\end{equation}
where ${\tau_{0i}}$ (${\tau_{Ii}}$) is the momentum relaxation time induced by intra-valley (inter-valley) scattering, and defined by
\begin{eqnarray}
\frac{1}{\tau_{0i}}&\equiv&\frac{2\pi}{\hbar}\underset{k'_x,k'_z}{\sum}|U^{1+i,0i}_{k_x,k_z;k'_x,k'_z}|^{2}\delta\left(E_{F}-E^{0i}_{k_z}\right), \label{time0}\\
\frac{1}{\tau_{Ii}}&\equiv&\frac{2\pi}{\hbar}\underset{k'_{x},k'_{z}}{\sum}|U_{k_{x},k_{z};k'_{x},k'_{z}}^{1+i,0\bar{i}}|^{2}
\delta\left(E_{F}-E^{0i}_{k_z}\right), \label{timeI}
\end{eqnarray}
where $\bar{i}=K'$ for $i=K$; while $\bar{i}=K$ for $i=K'$, and one element of $U$ matrix is shown here
\begin{equation}
U^{1+K,0K}_{k_x,k_z;k_x',k_z'} =\sin\frac{\theta_{1}^{k_{z}}}{2}I_{1,0}\left(\mathbf{R}_{i}\right).
\label{U1+K0K}
\end{equation}
Other elements of matrix $U$ can be found in Appendix \ref{U}. The details of the calculation of the momentum relaxation time is given in the Appendix \ref{xxtransport}. Then we obtain the total transverse magnetoconductivity
\begin{equation}
\sigma_{xx}=\sigma_{xx,\text{inter}}^{K}+\sigma_{xx,\text{intra}}^{K}+\sigma_{xx,\text{inter}}^{K^{\prime}}+\sigma_{xx,\text{intra}}^{K^{\prime}},
\label{totalsigmaxx}
\end{equation}
where
\begin{eqnarray}
\sigma_{xx,\text{inter}}^{K} &=& \frac{e^{2}}{h}\left(\frac{4\pi e^{2}}{\alpha\kappa_{0}}\right)^{2}\frac{\Lambda n_{\text{imp}}}{\hbar v_{F}}F_{1}^{K}(k_{z}){I_{1}}, \notag\\
\sigma_{xx,\text{intra}}^{K} &=& \frac{e^{2}}{h}\left(\frac{4\pi e^{2}}{\alpha\kappa_{0}}\right)^{2}\frac{\Lambda n_{\text{imp}}}{\hbar v_{F}}F_{2}^{K}(k_{z})\frac{I_{2}}{4},\notag\\
\sigma_{xx,\text{intra}}^{K^{\prime}}&=&\frac{e^{2}}{h}\left(\frac{4\pi e^{2}}{\alpha\kappa_{0}}\right)^{2}\frac{\Lambda n_{\text{imp}}}{\hbar v_{F}}F_{1}^{K^{\prime}}(k_{z})I_{3},\notag\\
\sigma_{xx,\text{inter}}^{K^{\prime}}&=&\frac{e^{2}}{h}\left(\frac{4\pi e^{2}}{\alpha\kappa_{0}}\right)^{2}\frac{\Lambda n_{\text{imp}}}{\hbar v_{F}}F_{2}^{K^{\prime}}(k_{z})\frac{I_{4}}{4}.
\label{sigmaInterIntra}
\end{eqnarray}.
We call $F$s form factors, and derive them in Appendix \ref{xxtransport}, as well as the expressions of $I_2$, $I_3$ and $I_4$.
%%%%%%%%%%%%%%%%%%%%%%%%%%%%%%%%%%%%%%%%%%%%%%%%%%%%%%%%%%%%%%%%%%%%%%%%%%%%%%%%%%%%%%%%%%%%%%%%%%%%%%%%%%%
\begin{figure}[tb]
\includegraphics[width=\linewidth]{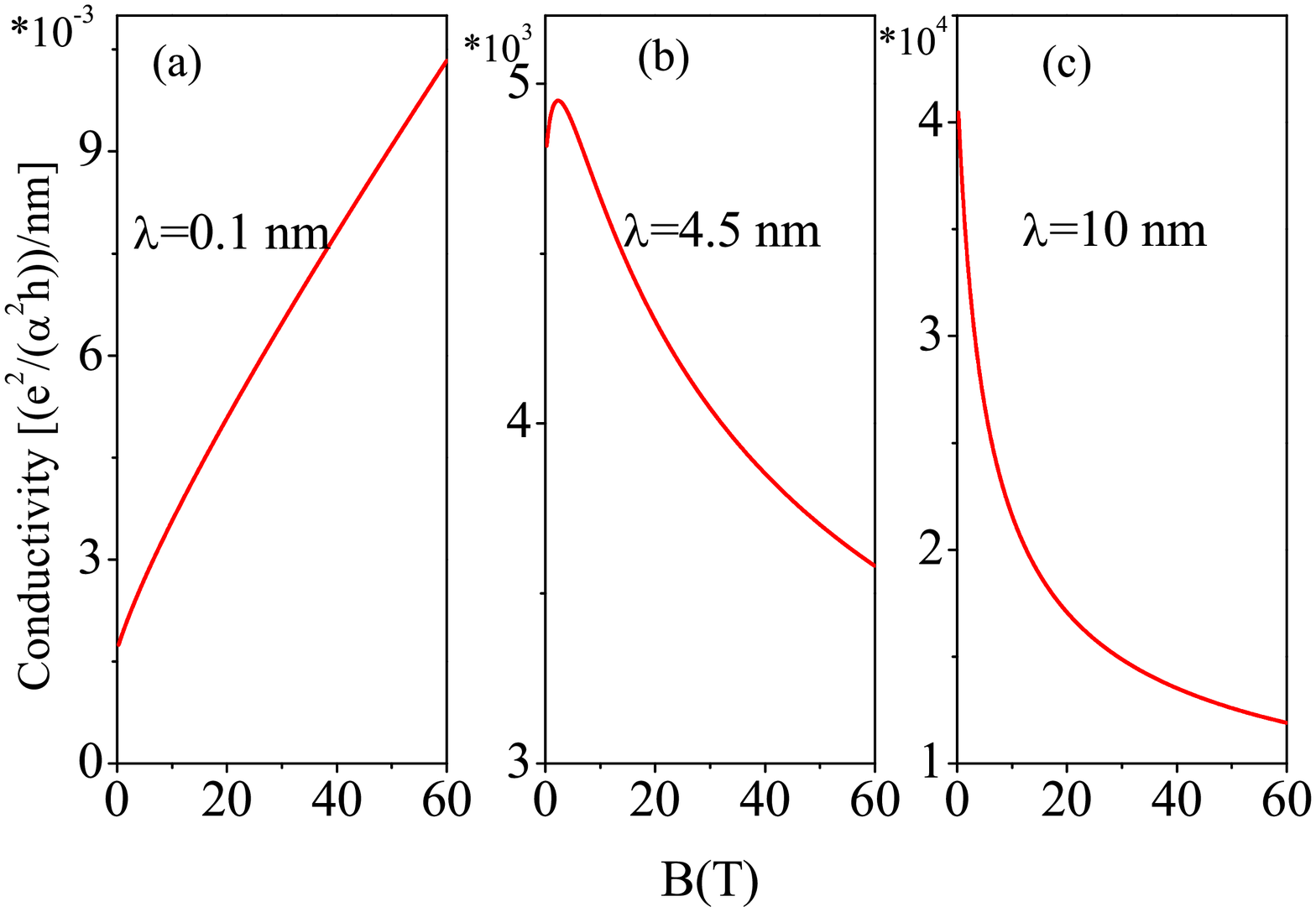}
%\vspace{-0.5cm}
\caption[]{The transverse magnetoconductivity, $\sigma_{xx}$, as a function of $B$ at different screening lengths for (a) $\lambda=0.1$ nm, (b) $\lambda=4.5$ nm, and (c) $\lambda=10$ nm, respectively. Here $k_{c}=0.1$ nm$^{-1}$.}\label{small}
\end{figure}
%%%%%%%%%%%%%%%%%%%%%%%%%%%%%%%%%%%%%%%%%%%%%%%%%%%%%%%%%%%%%%%%%%%%%%%%%%%%%%%%%%%%%%%%%%%%%%%%%%%%%%%%%%%
%%%%%%%%%%%%%%%%%%%%%%%%%%%%%%%%%%%%%%%%%%%%%%%%%%%%%%%%%%%%%%%%%%%%%%%%%%%%%%%%%%%%%%%%%%%%%%%%%%%%%%%%%%%
\begin{figure}[tb]
\includegraphics[width=\linewidth]{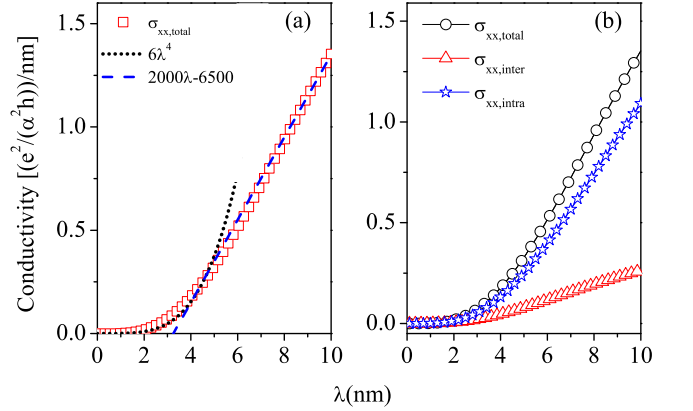}
%\vspace{-0.5cm}
\caption[]{The transverse magnetoconductivity, $\sigma_{xx}$, as a function of screening length $\lambda$ at $B=5T$. (a) is shown in logarithmic coordinates, (b) is in linear coordinates, respectively. Here $k_{c}=0.1$ nm$^{-1}$.}\label{sigmaxxfig}
\end{figure}
%%%%%%%%%%%%%%%%%%%%%%%%%%%%%%%%%%%%%%%%%%%%%%%%%%%%%%%%%%%%%%%%%%%%%%%%%%%%%%%%%%%%%%%%%%%%%%%%%%%%%%%%%%%

In Fig. \ref{small}, we plot the transverse magnetoconductivities as a function of $B$ at different screening length $\lambda$.
 The most important feature is that the magnitude of transverse magnetoconductivities are all enhanced sensitively in orders by increasing $\lambda$. This is just opposite to the behavior of the longitudinal magnetoconductivity which is suppressed by larger $\lambda$. We will come back this point later.
Another feature is about the $B$-dependence of the transverse magnetoconductivities. At small $\lambda$, $\sigma_{xx}$ increases with $B$ approximately linearly; while it is inversely proportional to $B$ at large $\lambda$. Therefore at some specific screening length, the two tendencies compete with each other, giving rise to a non-monotonic behavior of $\sigma_{xx}$ (see Fig. \ref{small}(b)).

In Fig. \ref{sigmaxxfig}, we show the $\lambda$-dependence of the transverse magnetoconductivities. It is seen from Fig. \ref{sigmaxxfig}(a) that the magnitude of the transverse magnetoconductivities is proportional to $\lambda^{4}$ for small $\lambda$, which is also supported by the $\lambda$ dependence of $I_{1},I_{2},I_{3}$ and $I_{4}$. This $\lambda^{4}$-dependence can also be derived analytically at $\lambda\rightarrow0$ limit. For a comparison, we use a linear function to fit the data when $\lambda\in$ (4 nm, 10 nm) in Fig. \ref{sigmaxxfig}(a). It is shown that they match very well. This linearity can not continue for much larger $\lambda$ but turns to a saturation in this case (not shown in the figure).  The inter-valley and intra-valley contributions are shown in Fig. \ref{sigmaxxfig}(b). We observe a similar $\lambda$-dependence to these two contributions as well.

The enhanced transverse magnetoconductivity with $\lambda$ can be understood by looking at the behavior of screened Coulomb disorder. The screened Coulomb potential can be reformulated $\phi(q)=\frac{4\pi e^2}{\alpha\kappa_{0} q^2}S(q,\lambda)$, where $S(q,\lambda)=\frac{(q\lambda)^2}{1+(q\lambda)^2}$ is a defined screening factor. Since we select the parameter $k_{c}=0.1$ nm$^{-1}$, for an inter-valley scattering $q$ is equal to $0.2$ nm$^{-1}$. $S(q,\lambda)$ varies monotonically from 0 to 1 with $q\lambda\in[0,\infty)$. For the fixed $q$=$0.2$ nm$^{-1}$, a larger $\lambda$ gives rise to a larger screening factor, the magnitude of screened Coulomb potential, the inter-valley scattering ($I_{1}$) and intra-valley scattering ($I_{2}$). In accordance with this increase, the transverse magnetoconductivity is enhanced because it is proportional to the relaxation times (in $I_{1}$ and $I_{2}$).

To study the relative weight of the inter-valley scattering and the intra-valley scattering, we define a ratio
\begin{eqnarray}
 R=\frac{\sigma_{xx,\text{inter}}}{\sigma_{xx,\text{intra}}},
 \label{R}
\end{eqnarray}
where $\sigma_{xx,\text{inter(intra)}}=\sigma^{K}_{xx,\text{inter(intra)}}+\sigma^{K'}_{xx,\text{inter(intra)}}$.
Therefore, $R=0$ reflects that there is only intra-valley scattering, and the inter-valley scattering is absent. When $R=\infty$, the situation is just reversed that there only exists inter-valley scattering. For finite $R$, both scattering channels contribute. When $R>1$, the inter-valley scattering dominates.

%%%%%%%%%%%%%%%%%%%%%%%%%%%%%%%%%%%%%%%%%%%%%%%%%%%%%%%%%%%%%%%%%%%%%%%%%%%%%%%%%%%%%%%%%%%%%%%%%%%%%%%%%%%
\begin{figure}[tb]
\includegraphics[width=1\linewidth]{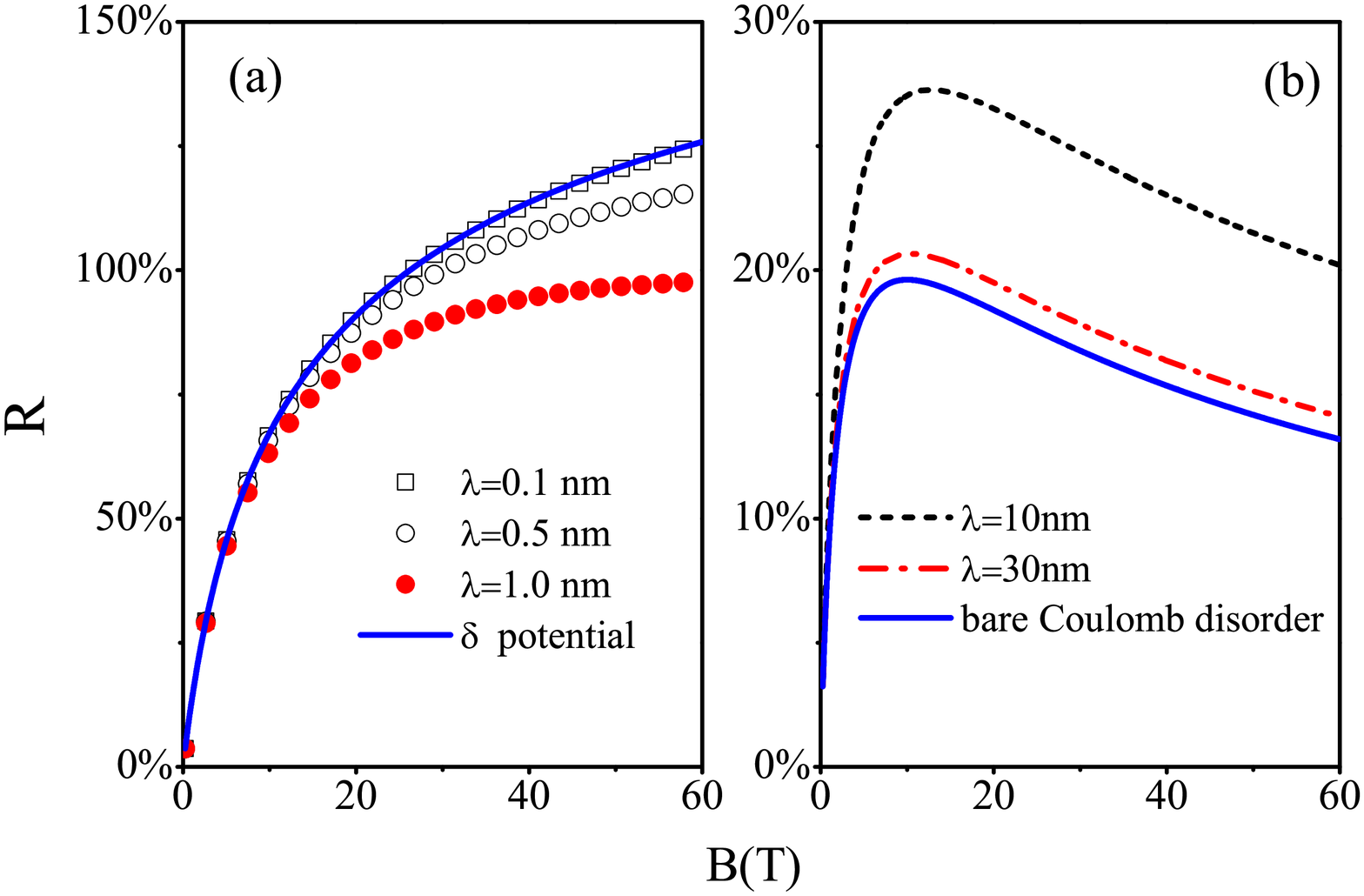}
\caption[]{The ratio $R$ versus $B$ at different screening lengths are studied. In (a), the red spots are for $\lambda=1.0$ nm; the open spots are for $\lambda=0.5$ nm, the open squares for $\lambda=0.1$ nm and the solid line represents the result of $\delta$-potential disorder, respectively. In (b), the short dashed line, dash-dot line and the solid line are for $\lambda=10$ nm, $\lambda=30$ nm and the bare Coulomb disorder, respectively. Here $k_{c}=0.1$ nm$^{-1}$.}\label{limit}
\end{figure}
%%%%%%%%%%%%%%%%%%%%%%%%%%%%%%%%%%%%%%%%%%%%%%%%%%%%%%%%%%%%%%%%%%%%%%%%%%%%%%%%%%%%%%%%%%%%%%%%%%%%%%%%%%%

In Fig. \ref{limit}(a), the dependence of $R$ on $B$ is shown for small screening lengths  $\lambda=0.1$ nm, $0.5$ nm, and $1$ nm, respectively. It is found that for small screening lengths the intra-valley scattering dominates under small $B$ and the ratio tends to a constant for large $B$. On the contrary, the inter-valley scattering dominates when $B$ is large. In Fig. \ref{limit}(b), the effect of large screening lengths on $R$ is considered. Although $R$ shows a non-monotonic behavior, $R$ is less than $1/3$ in the entire region of $B$. We may conclude that the intra-valley scattering dominates for a smoother disorder potential.

As $\lambda$ is reduced, the screened Coulomb potential is more local. Its limiting situation is a $\delta$-potential which is given by
\begin{eqnarray}
U\left(\mathbf{r}\right)=\underset{i}{\sum}u_{i}\delta\left(\mathbf{r}-\mathbf{R}_{i}\right).
\end{eqnarray}
In the same footing and by the virtue of the following two integrals,
\begin{eqnarray}
\tilde{I_{1}} &\equiv& \int\frac{dxdy}{\left(2\pi\right)^{2}}e^{-\left(x^{2}+y^{2}\right)/2}, \notag\\
%\end{eqnarray}
%\begin{eqnarray}
\tilde{I_{2}} &\equiv& \frac{l_{B}^{2}}{2}\int\frac{dxdy}{\left(2\pi\right)^{2}}\left(x^{2}+y^{2}\right)e^{-\left(x^{2}+y^{2}\right)/2},
\end{eqnarray}
we calculated the $\delta$-potential case. The results are shown in Fig. \ref{limit}(a).
It is seen that the calculated $R$ for the case of $\lambda=0.1$ nm and those of $\delta$-potential almost coincide with each other. The $\delta$-potential could serve as the lower boundary of screened Coulomb potential.

On the contrary, a large $\lambda$ indicates a more extended and smoother potential. Its limit is the bare Coulomb-potential disorder corresponding to $\lambda\rightarrow\infty$ in which there is no screening. We also calculated the consequences of this bare Coulomb disorder in Fig. \ref{limit}(b). It is seen that its results almost coincide with those of $\lambda=30$ nm. This confirms our expectation. The bare Coulomb disorder can serve as an upper boundary of the screened Coulomb disorder.
%

%%%%%%%%%%%%%%%%%%%%%%%%%%%%%%%%%%%%%%%%%%%%%%%%%%%%%%%%%%%%%%%%%%%%%%%%%%%%%%%%%%%%%%%%%%%%%%%%%%%%%%%%%%%
\begin{figure}[tb]
\includegraphics[width=\linewidth]{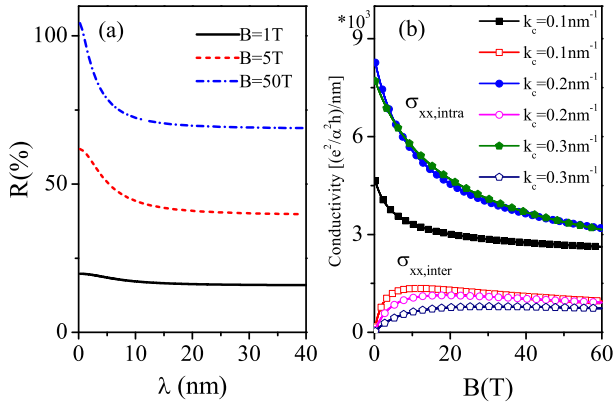}
%\vspace{-0.5cm}
\caption[]{In (a), the ratio $R$ varies with the screening length at different magnetic flux density, $k_{c}=0.1$ nm$^{-1}$. In (b), we show how $\sigma_{xx,\text{inter(intra)}}$ varies with $B$ at different $k_{c}$, and $\lambda=4.5$ nm. }\label{limitt}
\end{figure}
%%%%%%%%%%%%%%%%%%%%%%%%%%%%%%%%%%%%%%%%%%%%%%%%%%%%%%%%%%%%%%%%%%%%%%%%%%%%%%%%%%%%%%%%%%%%%%%%%%%%%%%%%%%

In Fig.~\ref{limitt}(a), the $\lambda$ dependence of the ratio $R$ is depicted at different $B$. It is clear that increasing the screening length has little effect on $R$ for small $B$. The ratio $R$ remains at low values (see $B=1$ T case). This means that the inter-valley contribution is tiny in this case, which is consistent with the results in Fig. \ref{limit}. The ratio $R$ only can be apparently large at a large $B$. We further note that $R$ tends to a constant after a turning $\lambda^{*}$. With increasing $B$, $\lambda^{*}$ becomes larger. Since in the laboratory range of $B$, $l_{B}$ is about several nanometers, i.e. at the order of $10$ nm, hence, when $\lambda>\lambda^{*}$, the impact of the increase of $\lambda$ becomes invalid.

To know more about the effect of inter-valley and intra-valley scattering, we show how $\sigma_{xx,\text{inter(intra)}}$ varies with $B$ at different $k_{c}$ in Fig. \ref{limitt}(b). For a fixed $k_{c}$, $\sigma_{xx,\text{inter(intra)}}$ increases (decreases) with $B$. It can be understood that the inter-valley scattering is facilitated at larger $B$ since the magnetic length is shortened and corresponding conductivity channels are enhanced for $\sigma_{xx}$. In contrast, $\sigma_{xx,\text{intra}}$ is depressed at larger $B$. For a fixed $B$, $\sigma_{xx,\text{inter}}$ is depressed with larger departure of the two nodes in the reciprocal space (larger $k_{c}$). This can be simply understood that a larger departure of the two nodes leads to harder transition between the two nodes, making smaller $\sigma_{xx,\text{inter}}$. We also witness a $1/B$-dependence of the intra-valley contribution to the transverse magnetoconductivities, which is consistent  with the results in  Refs. \onlinecite{Pesin,Klier} where only intra-valley scattering is considered. The inter-valley scattering can not be investigated in their one-node model. Therefore, the behavior of the inter-valley contribution to the transverse magnetoconductivities is new. And it is quite different from the intra-valley case. It is seen that the inter-valley scattering is as important as the intra-valley scattering. Without taking into account the inter-valley scattering, it is unlike to get a correct behavior of the total transverse magneto conductivity.

\subsection{Discussions and conclusions}
From recent experiments on Weyl semimetals, we observe several key facts. 1) The chemical potential is sample-dependent \cite{Long}; 2) The carrier density can be varied in several orders and the mobility can be different in one order \cite{Zhang:2015gwa,Long,arnold,Li-Hui}; 3) The longitudinal negative MR (related to $\sigma_{zz}$) corresponding to the chiral anomaly may not be observed in \textit{every} experiment \cite{Long}. From the first and second facts, we may conclude that there exist charged impurities in realistic samples and the density of the impurity is sample-dependent. According to our calculation, it is known that the longitudinal MR is very sensitive to the charged impurities. About the third fact, a reasonable explanation is given that the transverse resistivity (related to $\sigma_{xx}$) is usually quite larger than the longitudinal resistivity. In realistic measurements, it is hard to parallelize the magnetic field and electric field. Thus a perpendicular component of magnetic field $\mathbf{B}$ with respect to the electric field $\mathbf{E}$ may induce a large transverse MR which may overwhelm the small longitudinal MR. From our study, this may occur for the system with charged impurities of short screening length, for example $\lambda=0.1$ nm. We thus need to discuss how to observe the negative longitudinal MR in experiments. Due to the opposite effect of the screened Coulomb disorder on the $\sigma_{zz}$ and $\sigma_{xx}$, we may enhance the longitudinal resistivity and decrease the transverse resistivity by increasing the screening length. Eventually, the longitudinal resistivity may be larger than the transverse one. Therefore, the negative longitudinal MR may be easier to be observed in experiments in this case.

\section{Summary \label{Sec:summary}}
In this work, we study a two-node Weyl semimetal in the presence of magnetic fields. Screened Coulomb potential may be the main source of disorder in Weyl semimetal. We calculated the longitudinal and transverse magnetoconductivity with the screened Coulomb disorder.
We find that the longitudinal magnetoconductivity is positive in magnetic field, giving rises to a negative magnetoresistance, which agrees with the results in Ref. \onlinecite{Zhang}. In the transverse magnetoconductivity, we disclose a crossover from linear-B dependence in the small screening length limit to $1/B$ dependence in the large screening length limit.
We also note that the magnitude of the transverse magnetoconductivity is proportional to the quadruplicate of the screening length $\lambda$ for small $\lambda$. At small $\lambda$, the transverse magnetoconductivity $\sigma_{xx}$ increases with $B$; while it decreases with $B$ at large $\lambda$. At middle $\lambda$, $\sigma_{xx}$ may show a non-monotonic behavior. We define a ratio $R$ quantifying the inter-valley and intra-valley components. It is found that $R$ is quite small for a weak magnetic field, indicating the intra-valley scattering dominates. The inter-valley scattering dominates only in the presence of strong magnetic field and a smaller screening length.

Due to the opposite dependence of longitudinal and transverse magnetoconductivity on the screened Coulomb disorder, we discuss the possible explanation of recent experiments. Here we propose that increasing screening length may facilitate the observation of the negative longitudinal magnetoresistivity induced by the chiral anomaly. Moreover, introducing disorder may provide a helpful way to manipulate the valley degree in the realm of  valleytronics of the Weyl semimetal.

\section{SUPPLEMENTARY MATERIAL}
See supplementary material for the detailed calculations about the extra factor $\Lambda$, transport time, matrix U, derivation of Eq. (\ref{sigma-xx1}), and momentum relaxation time.

\section{Acknowledgment}
This work is supported by Hundred Talents Program of The Chinese Academy of Sciences and the NSFC (Grant No. 11674317). G. S. is supported in part by the MOST (Grant No. 2013CB933401), the NSFC (Grant No. 11474279) and the CAS (Grant No. XDB07010100). H. Z. Lu was supported by the National Key R \& D Program (Grant No. 2016YFA0301700), and the National Natural Science Foundation of China (Grant No. 11574127).

\begin{widetext}

\appendix
\section{The extra factor $\Lambda$ }\label{lambda}
When we calculate  the imaginary part of Green's function of the $n=0$ band, we need to deal with
\begin{eqnarray}
\frac{\frac{\hbar}{2\tau}}{[E_{F}-\hbar v_{F}k_{z}-\hbar v_{F}k_{c}]^{2}+(\frac{\hbar}{2\tau})^{2}},
\end{eqnarray}
In this work, we assume $v_{F}>0$. In this case,
we can write $a=\sqrt{\hbar v_{F}}$, $b=\sqrt{E_{F}-\hbar v_{F}k_{c}}$,
and $c=2\tau/\hbar$. A widely used approximation is that
\begin{eqnarray}
\int_{-\infty}^{\infty}dx\frac{1/c}{(a^{2}x^{2}-b^{2})^{2}+1/c^{2}}=\int_{-\infty}^{\infty}dx\pi\delta(a^{2}x^{2}-b^{2}).\nonumber
\end{eqnarray}
However, this leads to unphysical van Hove singularities at the band
edges. We correct this approximation with an extra factor $\Lambda$,
so that
\begin{eqnarray}
\int_{-\infty}^{\infty}dx\frac{1/c}{(a^{2}x^{2}-b^{2})^{2}+1/c^{2}}=\int_{-\infty}^{\infty}dx\pi\Lambda\delta(a^{2}x^{2}-b^{2}).\nonumber
\end{eqnarray}
The form of $\Lambda$ can be found as follows. First, the integral
can be found as
\begin{eqnarray}
\int_{-\infty}^{\infty}dx\frac{1/c}{(a^{2}x^{2}-b^{2})^{2}+1/c^{2}}=\frac{\sqrt{c}}{a}\frac{\pi}{\sqrt{2}}\frac{\sqrt{C^{2}+\sqrt{C^{4}+1}}}{\sqrt{C^{4}+1}},\nonumber \end{eqnarray}
where $C^{2}=b^{2}c$. On the other hand, using the property of the
delta function
\begin{eqnarray}
\pi\Lambda\int_{-\infty}^{\infty}dx\delta(a^{2}x^{2}-b^{2})=\pi\Lambda/ab.\label{f-integral}
\end{eqnarray}
So
\begin{eqnarray}
\Lambda=\frac{\sqrt{C^{4}+C^{2}\sqrt{C^{4}+1}}}{\sqrt{2(C^{4}+1)}}.\label{Eq:Lambda}
\end{eqnarray}
In the limit $C\gg1$, $\Lambda\rightarrow1$.

\section{transport time }\label{zztransport}
The transport time $\frac{\hbar}{\tau_{k_F}^{0K,\text{tr}}}$ is calculated as:
\begin{eqnarray}%\label{timer}
\frac{\hbar}{\tau_{k_F}^{0K,\text{tr}}}
= 2\pi \sum_{k_x',k_z'}
\langle |U^{0,0}_{k_x,k_F;k_x',k_z'}|^2 \rangle \Lambda\delta(E_F-E^{0K}_{k_z'})%\left(1-\frac{v^z_{0,k_z'}}{v_F}\right).
\times  \left(1-\frac{v^{zK}_{0,k_z'}}{v_F}\right).
\label{transtime}
\end{eqnarray}
%
%
%$U^{0,0}_{k_x,k_F;k_x',k_z'}$ is the scattering matrix elements.
To proceed the derivation, we define an integral for convenience
\begin{eqnarray}
I_{n,m}(\mathbf{R}_i)=\frac{1}{L_xL_z} \int d\mathbf{r} \varphi^{n*}_{k_x}(y) \varphi^{m}_{k_x'}(y) \Phi(\mathbf{r}-\mathbf{R}_i)
\times e^{i(k_x'-k_x)x+i(k_z'-k_z) z},
\label{I-def}
\end{eqnarray}
where
\begin{eqnarray}
\Phi(\mathbf{r}-\mathbf{R}_i)&=&
\int \frac{d\mathbf{q}}{(2\pi)^3} \Phi(\mathbf{q})
e^{i\mathbf{q}\cdot (\mathbf{r}-\mathbf{R}_i)}.
\end{eqnarray}
and $m,n$ are Landau level index, and
%
%The Fourier transformation of the potential $\Phi(\mathbf{r}-\mathbf{R}_i)$ is
\begin{eqnarray}
\Phi(\mathbf{q})=\int\frac{e^{2}}{\alpha\kappa_{0} r}e^{-\frac{r}{\lambda}}e^{-i\mathbf{q\cdot r}}d\mathbf{r}=\frac{4\pi e^{2}}{\alpha\kappa_{0}}\frac{1}{q^{2}+\frac{1}{\lambda^{2}}}.
\end{eqnarray}
According to $\int_{-\infty}^\infty dx e^{ikx}=2\pi\delta(k)$, we obtain
\begin{eqnarray}
I_{n,m}(\mathbf{R}_i)
= \int \frac{d^3\mathbf{q}}{2\pi L_xL_z} e^{-i\mathbf{q}\cdot \mathbf{R}_i}\Phi(\mathbf{q})\int dy \varphi^{n*}_{k_x}(y)\varphi^{m}_{k_x'}(y) \times e^{iq_y y}\delta(q_x+k_x'-k_x) \delta(q_z+k_z'-k_z).
\end{eqnarray}
We need to calculate
\begin{eqnarray}
&&\sum_{i,j}I_{n,m}(\mathbf{R}_i) I^*_{n',m'}(\mathbf{R}_j)=\sum_{i,j} \iint \frac{d^3\mathbf{q}d^3\mathbf{q}'}{(2\pi L_xL_z)^2}    \times\Phi(\mathbf{q})\Phi(\mathbf{q}')\delta(q_x+k_x'-k_x)
 \delta(q_z+k_z'-k_z) \notag\\
&\times& \iint dydy' \varphi^{n*}_{k_x}(y) \varphi^{m}_{k_x'}(y)
\varphi^{n'}_{k_x}(y') \varphi^{m'*}_{k_x'}(y')
e^{iq_{y}y-iq_{y'}y'}
e^{-i\mathbf{q}\cdot \mathbf{R}_i}e^{i\mathbf{q}'\cdot \mathbf{R}_j}\delta(q'_x+k_x'-k_x)
 \delta(q_z'+k_z'-k_z).
\end{eqnarray}
We define
\begin{eqnarray}
 \mathcal{I}_{n,m}^{n',m'}\equiv \langle \sum_{i,j}I_{n,m}(\mathbf{R}_i)
I^*_{n',m'}(\mathbf{R}_j)\rangle_{\text{imp}}.
\end{eqnarray}
Under the average of impurity configurations \cite{mahan}, we have
\begin{eqnarray}
\langle\underset{i,j}{\sum}e^{i\mathbf{q}_{1}\cdot\mathbf{R}_{i}}e^{i\mathbf{q}_{2}\cdot\mathbf{R}_{j}}\rangle_{imp}\approx N_{imp}\frac{\left(2\pi\right)^{3}}{V}\delta\left(\mathbf{q}_{1}+\mathbf{q}_{2}\right)
=n_{imp}\left(2\pi\right)^{3}\delta\left(\mathbf{q}_{1}+\mathbf{q}_{2}\right),
\end{eqnarray}
where $N_{imp}$ and $n_{imp}$ are the number and density of the
scattering centers, respectively.

Integrating over $\mathbf{q}'$, we obtain
\begin{eqnarray}
\mathcal{I}_{n,m}^{n',m'}&=&n_{imp}\int\frac{d^{3}\mathbf{q}}{2\pi L_{x}L_{z}}\Phi^{2}(\mathbf{q})\delta(q_{x}+k_{x}'-k_{x})
\delta(q_{z}+k_{z}'-k_{z})\int dy\varphi_{k_{x}}^{n*}(y)\varphi_{k_{x}'}^{m}(y)e^{iq_{y}y}\nonumber\\
&\times&\int dy'\varphi^{n'}_{k_x}(y')\varphi^{m'*}_{k_x'}(y')
e^{-iq_y y'},
\end{eqnarray}
where we have used the following equations:
\begin{eqnarray}
 \delta^2(q_x+k_x'-k_x) &=\frac{L_x}{2\pi}\delta(q_x+k_x'-k_x),\\
 \delta^2(q_z+k_z'-k_z) &=\frac{L_z}{2\pi} \delta(q_z+k_z'-k_z).
\end{eqnarray}
The scattering matrix elements between state $|0,k_x,k_z\rangle$ and state $|0,k_x',k_z'\rangle$ can be written as
\begin{eqnarray}
U^{0,0} \equiv&\langle 0,k_x,k_z |\Phi(\mathbf{r})|0,k_x',k_z'\rangle
=\int d\mathbf{r}\Psi_{0,k_{x},k_{z}}^{*}\left(\mathbf{r}\right)\Phi\left(\mathbf{r}\right)\Psi_{0,k_{x}^{\prime},k_{z}^{\prime}}\left(\mathbf{r}\right),
\end{eqnarray}
where $\Psi_{n,k_{x},k_{z}}\left(\mathbf{r}\right)$ are the wave functions
\begin{eqnarray}
\Psi_{n,k_{x},k_{z}}\left(\mathbf{r}\right) &=& \frac{1}{\sqrt{L_{x}L_{z}}}e^{ik_{x}x+ik_{z}z}\varphi_{k_{x}}^{n}\left(y\right),\nonumber\\
\varphi_{k_{x}}^{n}\left(y\right) &=& \langle y|n\rangle
=\frac{e^{-\left(y-y_{0}\right)^{2}/2l_{B}^{2}}}{\sqrt{n!2^{n}\sqrt{\pi}l_{B}}}\mathcal{H}_{n}\left(\frac{y-y_{0}}{l_{B}}\right),
\label{Hermite1}
\end{eqnarray}
where $\mathcal{H}_{n}$ are the Hermite polynomials, and $y_{0}=l^{2}_{B}k_{x}$.

Hence, $U^{0,0}=\underset{i}{\sum}I_{0,0}\left(\mathbf{R}_{i}\right)$ and $\langle|U^{0,0}|^2\rangle=\mathcal{I}_{0,0}^{0,0}$.
Then
\begin{eqnarray}
\mathcal{I}_{0,0}^{0,0}
 = n_{imp}\int\frac{d^{3}\mathbf{q}}{2\pi L_{x}L_{z}}\Phi^{2}(\mathbf{q})\delta(q_{x}+k_{x}'-k_{x})  \times  \delta(q_z+k_z'-k_z)W_{k_{x},k_{x}-q_{x},q_{y}}^{0,0},
\label{I0000}
\end{eqnarray}
where we define
\begin{eqnarray}
W_{k,k',q}^{m,n}=\Big| \int dy \varphi^{m*}_{k}(y)\varphi^{n}_{k'}(y)
e^{iqy} \Big|^2.
\label{W}
\end{eqnarray}
By means of the Gaussian integral, we have
\begin{eqnarray}
&& \int dy \varphi^{0*}_{k_x}(y)\varphi^{0}_{k_x-q_x}(y)
e^{iq_y y}
 = \frac{1}{  \sqrt{\pi}   } \int \frac{dy}{l_B} e^{-\frac{\left(y-l_B^2k_x\right)^2}{2l_B^2}}
e^{-\frac{\left[y-\left(k_x-q_x\right)l_B^2\right]^2}{2l_B^2}}
e^{iq_y y}
=  e^{-{l_B^2}\left(q_\perp^2-2 i q_x q_y + i 4 k_x q_y\right)/{4}},
\end{eqnarray}
where $q_\perp^2=q_x^2+q_y^2$, leading to
\begin{eqnarray}
\langle|U^{0,0}|^2\rangle
=n_{imp}\int\frac{d^{3}\mathbf{q}}{2\pi L_{x}L_{z}}\Phi^{2}(\mathbf{q})e^{-q_{\perp}^{2}l_{B}^{2}/2}
\times  \delta(q_x+k_x'-k_x) \delta(q_z+k_z'-k_z).\label{U-00}
\end{eqnarray}
Substituting Eq. (\ref{U-00}) into Eq. (\ref{transtime}), we obtain
\begin{eqnarray}
\frac{\hbar}{\tau_{k_F}^{\text{0K,tr}}}
= \frac{\Lambda n_{imp}}{\hbar v_F} \sum_{k_x',k_z'}
 \int  \frac{d^3\mathbf{q}}{L_xL_z } \Phi^2(\mathbf{q}) e^{- q_\perp^2l_B^2/2}
 \delta(q_x+k_x'-k_x)\delta(q_z+k_z'-k_z)
 \delta(k_z'+k_{F}-k_{c})\left(1-\frac{v^{zK}_{0,k_z'}}{v_F}\right).\label{transport time}
\end{eqnarray}
By changing the summations over $k_x',k_z'$ to integrals, Eq. (\ref{transport time}) becomes
\begin{eqnarray}
\frac{\hbar}{\tau_{k_F}^{\text{0K,tr}}}&=&\frac{\Lambda n_{imp}}{\hbar v_{F}}\int\frac{L_{z}}{2\pi}dk_{z}^{\prime}\delta(k_{z}'+k_{F}-k_{c})
\delta\left(q_{z}+k_{z}^{\prime}-k_{F}\right)\int\delta\left(q_{z}+k_{x}^{\prime}-k_{x}\right)\frac{dk_{x}^{\prime}L_{x}}{2\pi}\nonumber\\
&\times&\int\frac{d^{3}\mathbf{q}}{L_{x}L_{z}}\Phi^{2}(\mathbf{q})e^{-q_{\perp}^{2}l_{B}^{2}/2}
\left(1-\frac{v_{0,k_{z}'}^{zK}}{v_{F}}\right). \notag
\end{eqnarray}
Considering $1-\frac{v_{0,k_{z}'}^{zK}}{v_{F}}=2$, we have
\begin{eqnarray}
\frac{\hbar}{\tau_{k_F}^{\text{0K,tr}}}
=\frac{2\Lambda n_{imp}}{\hbar v_F}
\int \frac{dq_xdq_y}{(2\pi)^2 }    \Phi^2(q_x,q_y,2k_{F}-k_{c}) e^{-q_\perp^2l_B^2/2}
=\frac{2\Lambda n_{imp}}{\hbar v_{F}}\left(\frac{4\pi e^{2}}{\alpha\kappa_{0}}\right)^{2}l_{B}^{2}I_{1},
\end{eqnarray}
where
\begin{equation}
I_{1} \equiv \int\frac{dxdy}{\left(2\pi\right)^{2}}e^{-\left(x^{2}+y^{2}\right)/2}g(x,y),
\label{one}
\end{equation}
where $g(x,y)=\frac{1}{\left( x^{2}+y^{2}+l_{B}^{2}\left[\left(2k_{F}-k_{C}\right)^{2}+\frac{1}{\lambda^{2}}\right]\right)^{2}}$ is a defined function.
The summation over $k_x$ is limited by the Landau degeneracy and can be changed into an integral. Suppose that $L_x\gg \ell_B$, we get
\begin{eqnarray}
\sigma_{zz} & = & \frac{e^{2}}{h}\frac{v_F}{\pi L_{y}}\int_{-L_{y}/2l_{B}^{2}}^{L_{y}/2l_{B}^{2}}
dk_{x}\tau^{0K,\text{tr}}_{k_F}\Lambda .
\label{sigmazzAppendix}
\end{eqnarray}
then we have
\begin{eqnarray}
\sigma_{zz}
& = & \frac{e^{2}}{h}
\frac{(\hbar v_F)^2 }{2\pi n_{imp}\left(\frac{4\pi e^{2}}{\alpha\kappa_{0}}\right)^{2}I_{1}l_B^4}.\label{sigmazzfAppendix}
\end{eqnarray}

\section{Matrix U}\label{U}
From Eqs. (\ref{EigenstatesK}), (\ref{Eigen0K}), (\ref{EigenstatesKprime}), and (\ref{Eigen0Kprime}), we obtain
\begin{eqnarray}
\Psi_{n,k_{x},k_{z},\nu}\left(\mathbf{r}\right)&=&\langle\mathbf{r}|n,k_{x},k_{z},\nu\rangle= \frac{1}{\sqrt{L_{x}L_{z}}}e^{ik_{x}x+ik_{z}z}\varphi_{k_{x}}^{n}\left(y\right), \notag\\
\varphi_{k_{x}}^{n}\left(y\right) &=& \langle y|n\rangle=\frac{e^{-\left(y-y_{0}\right)^{2}/2l_{B}^{2}}}{\sqrt{n!2^{n}\sqrt{\pi}l_{B}}}\mathcal{H}_{n}\left(\frac{y-y_{0}}{l_{B}}\right),
\label{HermiteAppendix}
\end{eqnarray}
where $\nu=K,K^{\prime}$ for the valley index, and $n=0,1,2\dots$ for the Landau level index, respectively.

The matrix elements of scattering matrix $U$ are calculated in the following:
\begin{eqnarray}
U_{k_{x},k_{z};k_{x}^{\prime},k_{z}^{\prime}}^{1+K,0K}&=&\langle1,k_{x},k_{z},+K|\mathbf{r}\rangle\Phi\left(\mathbf{r-R}_{i}\right)\langle\mathbf{r}|0,k_{x}^{\prime},k_{z}^{\prime},K\rangle \nonumber\\
&=&\int d\mathbf{r}\left[\begin{array}{cc}
\cos\frac{\theta_{1}^{k_{z}}}{2}\langle0|y\rangle, & \sin\frac{\theta_{1}^{k_{z}}}{2}\langle1|y\rangle\end{array}\right]e^{-ik_{x}x-ik_{z}z}\Phi\left(\mathbf{r-R}_{i}\right)|\left[\begin{array}{c}
0\\
\langle y|0\rangle
\end{array}\right]e^{ik_{x}^{\prime}x+ik_{z}^{\prime}z} \nonumber\\
&=&\int\frac{d\mathbf{r}}{L_{x}l_{z}}\sin\frac{\theta_{1}^{k_{z}}}{2}\varphi_{k_{x}}^{1*}\left(y\right)\varphi_{k_{x}}^{0}\left(y\right)\Phi\left(\mathbf{r-R}_{i}\right)e^{i\left(k_{x}^{\prime}-k_{x}\right)x+i\left(k_{z}^{\prime}-k_{z}\right)z}\nonumber\\
&=&\sin\frac{\theta_{1}^{k_{z}}}{2}I_{1,0}\left(\mathbf{R}_{i}\right),
\end{eqnarray}

\begin{eqnarray}
U_{k_{x},k_{z};k_{x}^{\prime},k_{z}^{\prime}}^{1+K,0K^{\prime}}&=&\langle1,k_{x},k_{z},+K|\Phi\left(r\right)|0,k_{x}^{\prime},k_{z}^{\prime},K^{\prime}\rangle \nonumber\\
&=&\int d\mathbf{r}\left[\begin{array}{cc}
\cos\frac{\theta_{1}^{k_{z}}}{2}\langle0|y\rangle, & \sin\frac{\theta_{1}^{k_{z}}}{2}\langle1|y\rangle\end{array}\right]e^{-ik_{x}x-ik_{z}z}\Phi\left(\mathbf{r-R}_{i}\right)|\left[\begin{array}{c}
\langle y|0\rangle\\
0
\end{array}\right]e^{ik_{x}^{\prime}x+ik_{z}^{\prime}z} \nonumber\\
&=&\int\frac{d\mathbf{r}}{L_{x}l_{z}}\cos\frac{\theta_{1}^{k_{z}}}{2}\varphi_{k_{x}}^{0*}\left(y\right)\varphi_{k_{x}}^{0}\left(y\right)\Phi\left(\mathbf{r-R}_{i}\right)e^{i\left(k_{x}^{\prime}-k_{x}\right)x+i\left(k_{z}^{\prime}-k_{z}\right)z}\nonumber\\
&=&\cos\frac{\theta_{1}^{k_{z}}}{2}I_{0,0}\left(\mathbf{R}_{i}\right),
\end{eqnarray}

\begin{eqnarray}
U_{k_{x},k_{z};k_{x}^{\prime},k_{z}^{\prime}}^{1+K^{\prime},0K}&=&\langle1,k_{x},k_{z},+K^{\prime}|\Phi\left(r\right)|0,k_{x}^{\prime},k_{z}^{\prime},K\rangle\nonumber\\
&=&\int d\mathbf{r}\left[\begin{array}{cc}
\sin\frac{\theta_{1}^{k_{z}}}{2}\langle0|y\rangle, & -\cos\frac{\theta_{1}^{k_{z}}}{2}\langle1|y\rangle\end{array}\right]e^{-ik_{x}x-ik_{z}z}\Phi\left(\mathbf{r-R}_{i}\right)|\left[\begin{array}{c}
0\\
\langle y|0\rangle
\end{array}\right]e^{ik_{x}^{\prime}x+ik_{z}^{\prime}z} \nonumber\\
&=&-\int\frac{d\mathbf{r}}{L_{x}l_{z}}\cos\frac{\theta_{1}^{k_{z}}}{2}\varphi_{k_{x}}^{1*}\left(y\right)\varphi_{k_{x}}^{0}\left(y\right)\Phi\left(\mathbf{r-R}_{i}\right)e^{i\left(k_{x}^{\prime}-k_{x}\right)x+i\left(k_{z}^{\prime}-k_{z}\right)z}\nonumber\\
&=&-\cos\frac{\theta_{1}^{k_{z}}}{2}I_{1,0}\left(\mathbf{R}_{i}\right),
\end{eqnarray}

\begin{eqnarray}
U_{k_{x},k_{z};k_{x}^{\prime},k_{z}^{\prime}}^{1+K^{\prime},0K^{\prime}}&=&\langle1,k_{x},k_{z},+K^{\prime}|\Phi\left(r\right)|0,k_{x}^{\prime},k_{z}^{\prime},K^{\prime}\rangle \nonumber\\
&=&\int d\mathbf{r}\left[\begin{array}{cc}
\sin\frac{\theta_{1}^{k_{z}}}{2}\langle0|y\rangle, & -\cos\frac{\theta_{1}^{k_{z}}}{2}\langle1|y\rangle\end{array}\right]e^{-ik_{x}x-ik_{z}z}\Phi\left(\mathbf{r-R}_{i}\right)|\left[\begin{array}{c}
\langle y|0\rangle\\
0
\end{array}\right]e^{ik_{x}^{\prime}x+ik_{z}^{\prime}z} \nonumber\\
&=&\int\frac{d\mathbf{r}}{L_{x}l_{z}}\sin\frac{\theta_{1}^{k_{z}}}{2}\varphi_{k_{x}}^{0*}\left(y\right)\varphi_{k_{x}}^{0}\left(y\right)\Phi\left(\mathbf{r-R}_{i}\right)e^{i\left(k_{x}^{\prime}-k_{x}\right)x+i\left(k_{z}^{\prime}-k_{z}\right)z} \nonumber\\
&=&\sin\frac{\theta_{1}^{k_{z}}}{2}I_{0,0}\left(\mathbf{R}_{i}\right).
\end{eqnarray}
$I_{n,m}(\mathbf{R}_i)$ is defined as
\begin{eqnarray}
I_{n,m}(\mathbf{R}_i)&=&\frac{1}{L_xL_z} \int d\mathbf{r} \varphi^{n*}_{k_x}(y) \varphi^{m}_{k_x'}(y) \Phi(\mathbf{r}-\mathbf{R}_i) \notag \\
& \times & e^{i(k_x'-k_x)x+i(k_z'-k_z) z},
\label{I-def1Appendix}
\end{eqnarray}
where $m,n$ represent the Landau level indexes.

\section{Derivation of Eq. (\ref{sigma-xx1})} \label{matrixformsigmaxx}
To derive $\sigma_{xx}$, we need extend the Green's functions and velocity operators in Landau-band space. However we only work in vicinity of Weyl nodes, therefore the probability transitions induced by the impurities only involve $0$-th band and $n=1+$ band. Thus the band space is reduced. The matrixes of the Green's functions are diagonal. The matrix elements of the velocity operator represent different scattering processes. In this sense, we have
\begin{eqnarray}\label{sigma-xx1Appendix}
\sigma_{xx}&=&\frac{e^{2}\hbar}{2\pi V}\underset{k_{x},k_{z}}{\sum}\text{Tr}\left\{\left(\begin{array}{cccc}
G_{0K}^{R} & 0 & 0 & 0\\
0 & G_{0K^{\prime}}^{R} & 0 & 0\\
0 & 0 & G_{1+K}^{R} & 0\\
0 & 0 & 0 & G_{1+K^{\prime}}^{R}
\end{array}\right)\left(\begin{array}{cccc}
0 & 0 & v_{0K,1+K}^{x} & v_{0K,1+K^{\prime}}^{x}\\
0 & 0 & v_{0K^{\prime},1+K}^{x} & v_{0K^{\prime},1+K^{\prime}}^{x}\\
v_{1+K,0K}^{x} & v_{1+K,0K^{\prime}}^{x} & 0 & 0\\
v_{1+K^{\prime},0K}^{x} & v_{1+K^{\prime},0K^{\prime}}^{x} & 0 & 0
\end{array}\right) \right.\nonumber\\
&\times&\left.\left(\begin{array}{cccc}
G_{0K}^{A} & 0 & 0 & 0\\
0 & G_{0K^{\prime}}^{A} & 0 & 0\\
0 & 0 & G_{1+K}^{A} & 0\\
0 & 0 & 0 & G_{1+K^{\prime}}^{A}
\end{array}\right)\left(\begin{array}{cccc}
0 & 0 & v_{0K,1+K}^{x} & v_{0K,1+K^{\prime}}^{x}\\
0 & 0 & v_{0K^{\prime},1+K}^{x} & v_{0K^{\prime},1+K^{\prime}}^{x}\\
v_{1+K,0K}^{x} & v_{1+K,0K^{\prime}}^{x} & 0 & 0\\
v_{1+K^{\prime},0K}^{x} & v_{1+K^{\prime},0K^{\prime}}^{x} & 0 & 0
\end{array}\right)\right\}\label{sigmaxxMatrix}\notag\\
&=&\frac{e^{2}\hbar}{\pi V}\underset{k_{x,}k_{z},i,i'}{\sum}\Re\left(G_{0i}^{R}v_{0i,1+i'}^{x}G_{1+i'}^{A}v_{1+i',0i}^{x}\right),
\label{matrixSigmaxx}
\end{eqnarray}
where $i,i'$ run over $K$ and $K'$ indexes.

\section{momentum relaxation time }\label{xxtransport}
With the aid of the integration $\mathcal{I}$ defined above, we obtain
\begin{eqnarray}
|U^{1+K,0K}_{k_x,k_z;k_x',k_z'}|^{2} &=& [\Theta^{k_{z}}_{1-}/2]\mathcal{I}_{1,0}^{1,0}, \notag\\
|U^{1+K,0K^{\prime}}_{k_x,k_z;k_x',k_z'}|^{2}&=&[\Theta^{k_{z}}_{1+}/2]\mathcal{I}_{0,0}^{0,0},  \notag\\
|U^{1+K^{\prime},0K^{\prime}}_{k_x,k_z;k_x',k_z'}|^{2}&=&[\Theta^{k_{z}}_{1-}/2]\mathcal{I}_{0,0}^{0,0}, \notag\\
|U^{1+K^{\prime},0K}_{k_x,k_z;k_x',k_z'}|^{2}&=&[\Theta^{k_{z}}_{1+}/2]\mathcal{I}_{1,0}^{1,0},
\end{eqnarray}
where $\Theta^{k_{z}}_{1\pm}=1\pm\cos\theta_{1}^{k_{z}}$, $\mathcal{I}_{0,0}^{0,0}$ is given in Eq. (\ref{I0000}), and
$\mathcal{I}_{1,0}^{1,0}$ takes the form of
\begin{eqnarray}
 \mathcal{I}_{1,0}^{1,0}
= \frac{n_{imp} l_{B}^{2}}{2}\int \frac{d^3\mathbf{q}}{2\pi L_xL_z } \Phi^2(\mathbf{q})W_{k_x,k'_x,q_y}^{1,0}\times \delta(q_x+k_x'-k_x)
 \delta(q_z+k_z'-k_z),
\end{eqnarray}
where $W$ is defined in Eq. (\ref{W}).
From Eqs. (\ref{time0}) and (\ref{timeI}), we obtain
\begin{eqnarray}
\frac{\hbar}{\tau_{0K}}&=&\Theta^{k_{z}}_{1-}\frac{l_{B}^{2}}{4}\frac{\Lambda n_{imp}}{\hbar v_{F}}\left(\frac{4\pi e^{2}}{\alpha\kappa_{0}}\right)^{2}I_{2},\\
\frac{\hbar}{\tau_{0K^{\prime}}}&=&\Theta^{k_{z}}_{1-}l_{B}^{2}\frac{\Lambda n_{imp}}{\hbar v_{F}}\left(\frac{4\pi e^{2}}{\alpha\kappa_{0}}\right)^{2}I_{3},\\
\frac{\hbar}{\tau_{IK}}&=&\Theta^{k_{z}}_{1+}l_{B}^{2}\frac{\Lambda n_{imp}}{\hbar v_{F}}\left(\frac{4\pi e^{2}}{\alpha\kappa_{0}}\right)^{2}I_{1},\\
\frac{\hbar}{\tau_{IK^{\prime}}}&=&\Theta^{k_{z}}_{1+}\frac{l_{B}^{2}}{4}\frac{\Lambda n_{imp}}{\hbar v_{F}}\left(\frac{4\pi e^{2}}{\alpha\kappa_{0}}\right)^{2}I_{4},
\end{eqnarray}
where
\begin{equation}
I_{2} \equiv \int\frac{dxdy}{\left(2\pi\right)^{2}}e^{-\left(x^{2}+y^{2}\right)/2} \left(x^{2}+y^{2}\right)g(x,y).
\label{two}
\end{equation}
\begin{equation}
I_{3} \equiv \int\frac{dxdy}{\left(2\pi\right)^{2}}e^{-\left(x^{2}+y^{2}\right)/2}h(x,y),
\label{three}
\end{equation}
\begin{equation}
I_{4} \equiv \int\frac{dxdy}{\left(2\pi\right)^{2}}e^{-\left(x^{2}+y^{2}\right)/2} \left(x^{2}+y^{2}\right)h(x,y),
\label{fourAppendix}
\end{equation}
where $h(x,y)=\frac{1}{\left( x^{2}+y^{2}+l_{B}^{2}\left[k_{C}^{2}+\frac{1}{\lambda^{2}}\right]\right)^{2}}$ is a defined function.%{\color{blue}{Similar as the case of $g(x,y)$, a $\lambda^{4}$-dependence of $h(x,y)$ can also be obtained.}}

The total transverse magneto-conductivity can be expressed as
\begin{equation}
\sigma_{xx}=\sigma_{xx,\text{inter}}^{K}+\sigma_{xx,\text{intra}}^{K}+\sigma_{xx,\text{inter}}^{K^{\prime}}+\sigma_{xx,\text{intra}}^{K^{\prime}},
\label{totalsigmaxxAppendix}
\end{equation}
where
\begin{eqnarray}
\sigma_{xx,\text{inter}}^{K} &=& \frac{e^{2}}{h}\left(\frac{4\pi e^{2}}{\alpha\kappa_{0}}\right)^{2}\frac{\Lambda n_{imp}}{\hbar v_{F}}F_{1}^{K}(k_{z}){I_{1}}, \notag\\
\sigma_{xx,\text{intra}}^{K} &=& \frac{e^{2}}{h}\left(\frac{4\pi e^{2}}{\alpha\kappa_{0}}\right)^{2}\frac{\Lambda n_{imp}}{\hbar v_{F}}F_{2}^{K}(k_{z})\frac{I_{2}}{4},\notag\\
\sigma_{xx,\text{intra}}^{K^{\prime}}&=&\frac{e^{2}}{h}\left(\frac{4\pi e^{2}}{\alpha\kappa_{0}}\right)^{2}\frac{\Lambda n_{imp}}{\hbar v_{F}}F_{1}^{K^{\prime}}(k_{z})I_{3},\notag\\
\sigma_{xx,\text{inter}}^{K^{\prime}}&=&\frac{e^{2}}{h}\left(\frac{4\pi e^{2}}{\alpha\kappa_{0}}\right)^{2}\frac{\Lambda n_{imp}}{\hbar v_{F}}F_{2}^{K^{\prime}}(k_{z})\frac{I_{4}}{4}.
\label{sigmaInterIntraAppendix}
\end{eqnarray}

Four form-factors are defined as follows
\begin{eqnarray}
F_{1}^{K}\left(k_{z}\right) &\equiv& 2\frac{|\hbar v_{0K^{\prime},1+K}^{x}|^{2}}{\left(E_{F}-E_{+K}^{1} \right)^{2}}\cos^{2}\left(\frac{\theta_{1}^{k_{z}}}{2}\right), \nonumber\\
F_{2}^{K}\left(k_{z}\right) &\equiv& 2\frac{|\hbar v_{0K,1+K}^{x}|^{2}}{\left(E_{F}-E_{+K}^{1} \right)^{2}}\sin^{2}\left(\frac{\theta_{1}^{k_{z}}}{2}\right),\nonumber\\
F_{1}^{K^{\prime}}\left(k_{z}\right) &\equiv& 2\frac{|\hbar v_{0K^{\prime},1+K^{\prime}}^{x}|^{2}}{\left(E_{F}-E_{+K^{\prime}}^{1} \right)^{2}}\sin^{2}\left(\frac{\theta_{1}^{k_{z}}}{2}\right),\nonumber\\
F_{2}^{K^{\prime}}\left(k_{z}\right) &\equiv& 2\frac{|\hbar v_{0K,1+K^{\prime}}^{x}|^{2}}{\left(E_{F}-E_{+K^{\prime}}^{1} \right)^{2}}\cos^{2}\left(\frac{\theta_{1}^{k_{z}}}{2}\right).
\end{eqnarray}
%The first term of Eq. (\ref{sigmaxxF1F2}) stems from the inter-valley scattering and the second term is induced by the intra-valley scattering, respectively.

When the Fermi energy is located in the vicinity of the Weyl nodes, we get
\begin{eqnarray}
F_{1}^{K}\left(k_{c}\right)=\frac{1}{2}\frac{\left(k_{c}+\sqrt{k_{c}^{2}+\frac{2}{l_{B}^{2}}}\right)^{2}}{\left(k_{c}^{2}+\frac{2}{l_{B}^{2}}\right)\left(k_{c}-\sqrt{k_{c}^{2}+\frac{2}{l_{B}^{2}}}\right)^{2}},\label{F1k}\\
F_{2}^{K}\left(k_{c}\right)=\frac{1}{2}\frac{\frac{2}{l_{B}^{2}}}{\left(k_{c}^{2}+\frac{2}{l_{B}^{2}}\right)\left(k_{c}-\sqrt{k_{c}^{2}+\frac{2}{l_{B}^{2}}}\right)^{2}},
\label{F2k}\\
F_{1}^{K^{\prime}}\left(-k_{c}\right)=\frac{1}{2}\frac{\left(k_{c}+\sqrt{k_{c}^{2}+\frac{2}{l_{B}^{2}}}\right)^{2}}{\left(k_{c}^{2}+\frac{2}{l_{B}^{2}}\right)\left(k_{c}-\sqrt{k_{c}^{2}+\frac{2}{l_{B}^{2}}}\right)^{2}}, \label{F1kprime}\\
F_{2}^{K^{\prime}}\left(-k_{c}\right)=\frac{1}{2}\frac{\frac{2}{l_{B}^{2}}}{\left(k_{c}^{2}+\frac{2}{l_{B}^{2}}\right)\left(k_{c}-\sqrt{k_{c}^{2}+\frac{2}{l_{B}^{2}}}\right)^{2}}\label{F2kprime}.
\end{eqnarray}

\end{widetext}

%

% ---------------------------------------------------------------------------
%-----------------------------------------------------------------------
%   END DOCUMENT
%-----------------------------------------------------------------------
\end{document}